\documentclass[twocolumn]{aastex63}

\usepackage{amssymb}
\usepackage{natbib}
\usepackage{amsmath}
\usepackage{comment}
\usepackage{amsthm}
\usepackage{multirow}
\usepackage[caption=false]{subfig}

\submitjournal{ApJL}

\begin{document}

\title{Golden galactic binaries for LISA: mass-transferring white dwarf black hole binaries}



\author[0000-0002-8751-9889]{Laura Sberna}
\affiliation{Perimeter Institute, 31 Caroline St N, Waterloo, ON N2L 2Y5, Canada}

\author[0000-0002-2685-1538]{Alexandre Toubiana}
\affiliation{APC, AstroParticule et Cosmologie, Universit\'e Paris Diderot,
	CNRS/IN2P3, CEA/Irfu, Observatoire de Paris, \\ Sorbonne Paris Cit\'e,
	10, rue Alice Domon et L\'eonie Duquet 75205 Paris Cedex 13, France}
\affiliation{Institut d'Astrophysique de Paris, CNRS \& Sorbonne
	Universit\'es, UMR 7095, 98 bis bd Arago, 75014 Paris, France}

\author{M. Coleman Miller}
\affiliation{Department of Astronomy { and Joint Space-Science Institute}, University of Maryland College Park, MD 20742-2421}

\begin{abstract}
We study the evolution and gravitational wave emission of white dwarf --- black hole accreting binaries with a semi-analytical model. These systems will evolve across the mHz gravitational wave frequency band and potentially be detected by the Laser Interferometer Space Antenna (LISA). We identify new universal relations for this class of binaries, which relate the component masses to the gravitational wave frequency and its first derivative. Combined with the high precision measurements possible with LISA, these relations could allow us to infer the component masses and the luminosity distance of the source. LISA has therefore the potential to detect and characterize a virtually unexplored binary population.
\end{abstract}


\section{Introduction} \label{sec:intro}

Galactic compact binaries will form a stochastic foreground signal that, { from ${\rm few}\times 10^{-4}~{\rm Hz}$ to ${\rm few}\times 10^{-3}~{\rm Hz}$, will dominate over} the instrumental noise of the Laser Interferometer Space Antenna (LISA) \citep{AHO17}, a gravitational wave (GW) space-borne experiment scheduled for { launch in} 2034~\citep{2001A&A...365..491N,2001A&A...368..939N, 2001A&A...375..890N, Liu_2010, 2010ApJ...717.1006R,Yu2010}. In addition to this foreground, LISA is expected to individually resolve $\sim 10^4$ compact binaries~(\cite{Nelemans:2003ha,Kremer:2017xrg,Korol17,2019MNRAS.490.5888L,Breivik_2020}). Among Galactic binaries, double white dwarfs (DWDs) are predicted to be the most numerous source. These binaries will be observed both in the mass-accreting and in the detached phase and could be targeted by other surveys in the electromagnetic band, such as Gaia~\citep{Breivik_2018}. The detection of such a large number and wide range of white dwarf (WD) binaries will allow the Milky Way {to be mapped} \citep{PhysRevD.86.124032, KorolBarausse, Breivik:2019oar}, to explore Milky Way satellites (\cite{Korol:2020lpq}, \cite{Roebber:2020hso}), measure the influence of tidal couplings (\cite{Fuller_2012},\cite{Shah:2015cua}), test binary population models~\citep{Toonen} and even test General Relativity~\citep{Littenberg_2019}.

Little attention has so far been devoted to another class of galactic binaries: accreting white dwarf --- black hole binaries (WDBH) {(see however~\cite{2012AA...537A.104V})}. Population studies predict that tens of thousand of mass-transferring WDBHs could form in the Milky Way (see e.g.~\cite{Hurley_2002}, \cite{Yungelson}), but the rates are still uncertain by more than an order of magnitude. The expectation is that binaries containing a black hole (BH) will be subdominant in the range of frequencies relevant for LISA ($ 0.1-1 \, {\rm mHz} $, see e.g.~\cite{2001AA375890N}). 
Although~\cite{Breivik_2020} suggest that LISA might not see any detached WDBHs in the Galaxy,~\cite{Kremer:2018tzm} find that a few events could be possible if we account for binary interactions in Galactic globular clusters. Overall, these sources are often discarded in BH population synthesis simulations (e.g.~\cite{2018MNRAS.480.2704L}) and further investigations are needed to predict the rate of their mass-transferring phase. 

There are no confirmed observations of WDBH binaries from electromagnetic surveys, although these binaries, like other mass-transferring systems, are expected to emit across a broad spectrum and have even been suggested to produce gamma-ray bursts~\citep{2018MNRAS.475L.101D}. The X-ray binary X-9, in the globular cluster 47 Tucanae, { might} host a WD and a BH~(\cite{10.1093/mnras/stv1869}, \cite{Tudor_2018}), but the system is also consistent with a neutron star accretor. Other candidates include XMMUJ$122939.7+075333$ in a globular cluster of the Virgo Galaxy NGC 4472~\citep{2007Natur.445..183M}. 
LISA will thus provide a complementary investigation of this elusive population and might be the first observatory to confirm their existence.

In this Letter, we demonstrate that a LISA observation of a WDBH binary would { reveal} the component masses and the luminosity distance of the system. In Sec.~\ref{sec:equations}, we describe our semi-analytical model to evolve WDBH binaries. We then determine two universal relations followed by these binaries in their evolution: one common to binaries with a WD (Helium) donor~(e.g. \cite{2005ASPC..330...27N}, \cite{Breivik_2018}) and one, first appearing in this work, applicable to accreting binaries with small tidal interactions. In Sec.~\ref{sec:LISA} we use these relations to infer the WD mass $M_{\rm WD}$, the BH mass $M_{\rm BH}$, and the luminosity distance $D_L$ from a LISA measurement of the GW amplitude { and the} frequency $ f $ and its first derivative $ \dot{f} $. 

\section{Evolution of mass transferring WDBH binaries}\label{sec:equations}
We consider WDBH binaries on a circular orbit with separation $a$. We model their evolution from the onset of mass transfer, when the WD overfills its Roche lobe. Our treatment follows that of~\cite{Marsh:2003rd}, with appropriate adjustments for the BH component. We use the zero-temperature mass-radius relation of~\cite{Verb&Rap} for the WD\footnote{Note that the accretion disk surrounding the BH can heat the WD. We will discuss this caveat further in the conclusions.}.
We define the total mass $ M= M_{\rm BH}+M_{\rm WD} $ and the mass ratio $ q= M_{\rm WD}/M_{\rm BH} \le 1  $.

\subsection{Mass transfer}

The overfill factor indicates by how much the donor overfills its Roche lobe, $\Delta=R_{\rm WD}-R_L$. 
Mass transfer occurs when $\Delta>0$ and increases monotonically with the overfill. We use the adiabatic approximation of~\cite{Marsh:2003rd} (see also \cite{1984ApJ277355W}):
\begin{align}\label{eq:M2eq}
\dot{M}_{\rm WD}  =-F(M_{\rm BH},M_{\rm WD},a,R_{\rm WD})\Delta^3 \, . 
\end{align}
See~\cite{Marsh:2003rd} for the definition of $ F $. 
%
%
%
We assume an accretion disk forms around the BH and that matter is transferred from the innermost stable circular orbit (ISCO) at a radius $R_{\rm ISCO}$~\citep{Chandrasekhar1984}.
We account for the limited efficiency of the BH to accrete by setting:
\begin{equation}\label{eq:M1eq}
\dot{M}_{\rm BH} = {\rm min} \left(-\dot{M}_{WD} \ \epsilon_{\rm ISCO},\dot{M}_{\rm Edd}(M_{\rm BH})\right),
\end{equation}
where $ \dot{M}_{\rm Edd} = 2.2 \times 10^{-8} M_{\rm BH}  \ {\rm year}^{-1}$ is the Eddington accretion rate and $ \epsilon_{\rm ISCO}$ is the specific mass-energy at the ISCO~\citep{Chandrasekhar1984}. Therefore mass is not necessarily conserved, accounting for possible loss through winds. 

\subsection{Orbital separation}
{We assume that the variation of total angular momentum is due to GW emission and loss of matter:}
\begin{equation}
\dot{J}_{\rm orb}+\dot{J}_{\rm BH}+\dot{J}_{\rm WD}=-\dot{J}_{\rm GW} -\dot{J}_{\rm loss}\label{ev_jorb},
\end{equation}
with $ \dot{J}_{\rm GW} = \frac{32}{5} \frac{G^3}{c^5} \frac{M_{\rm BH} M_{\rm WD} M}{a^4 }  \, J_{\rm orb} $. Following~\cite{2012AA...537A.104V}, we assume isotropic re-emission and take $\dot{J}_{\rm loss}=-q\frac{\dot{M}}{M}J_{\rm orb}$.  
We neglect the angular momentum of the accretion disk surrounding the BH,
assuming that $ M_{\rm disk} \ll M_{\rm BH}$ throughout the evolution. 

We assume that the WD is tidally locked. This is justified in low-mass-ratio systems such as WDBH binaries, since the synchronization time-scale decreases as the mass ratio squared, $ \tau_{\rm sync} \sim q^2$~\citep{10.1093/mnras/207.3.433}. Moreover, disk accretion can also contribute to synchronizing the star rotation with the orbit~\citep{1977A&A....57..383Z}.
The angular momentum of the donor can then be written as $J_{\rm WD}=I_{\rm WD} \Omega $, $\Omega$ being the orbital angular frequency and $I_{\rm WD}=k M_{\rm WD}R_{\rm WD}^2$ the momentum of inertia of the donor. The factor $ k $ is a function of the WD mass, for which we use the fit provided in~\cite{Marsh:2003rd}. Using Kepler's law, the variation in angular momentum of the donor is:
\begin{align}\label{eq:donorJ}
\dot{J}_{\rm WD}= I_{\rm WD} \Omega \Bigg( &\lambda \frac{\dot{M}_{\rm WD}}{M_{\rm WD}} -\frac{3}{2}\frac{\dot{a}}{a} \nonumber \\
&+ \frac{\dot{M}_{\rm BH}+ \dot{M}_{\rm WD}}{M_{\rm WD}} \frac{1}{2\left(1+1/q\right)}\Bigg ) , 
\end{align}
where $\lambda=1+2 \frac{{\rm d} \log R_{\rm WD}}{{\rm d} \log  M_{\rm WD}}+\frac{{\rm d} \log k}{{\rm d} \log  M_{\rm WD}}$. Note that the variation of the donor angular momentum was not included in the treatment of~\cite{Marsh:2003rd}.

We assume no tidal torque acts on the BH, so its angular momentum varies only as a result of the matter accreted at $R_{\rm ISCO}$,
\begin{equation}
\dot{J}_{\rm BH}= j_{\rm ISCO} \dot{M}_{\rm BH} \label{eq:spin1} ,
\end{equation}
where $j_{\rm ISCO}$ is the specific angular momentum at the ISCO~\citep{Chandrasekhar1984}. 
%
%
%
%
%
%
%
{ In Appendix~\ref{app:sep} we derive the resulting equation for the binary separation as a function of time.}
\begin{figure}[t]
	\centering
	\includegraphics[width=0.99\linewidth]{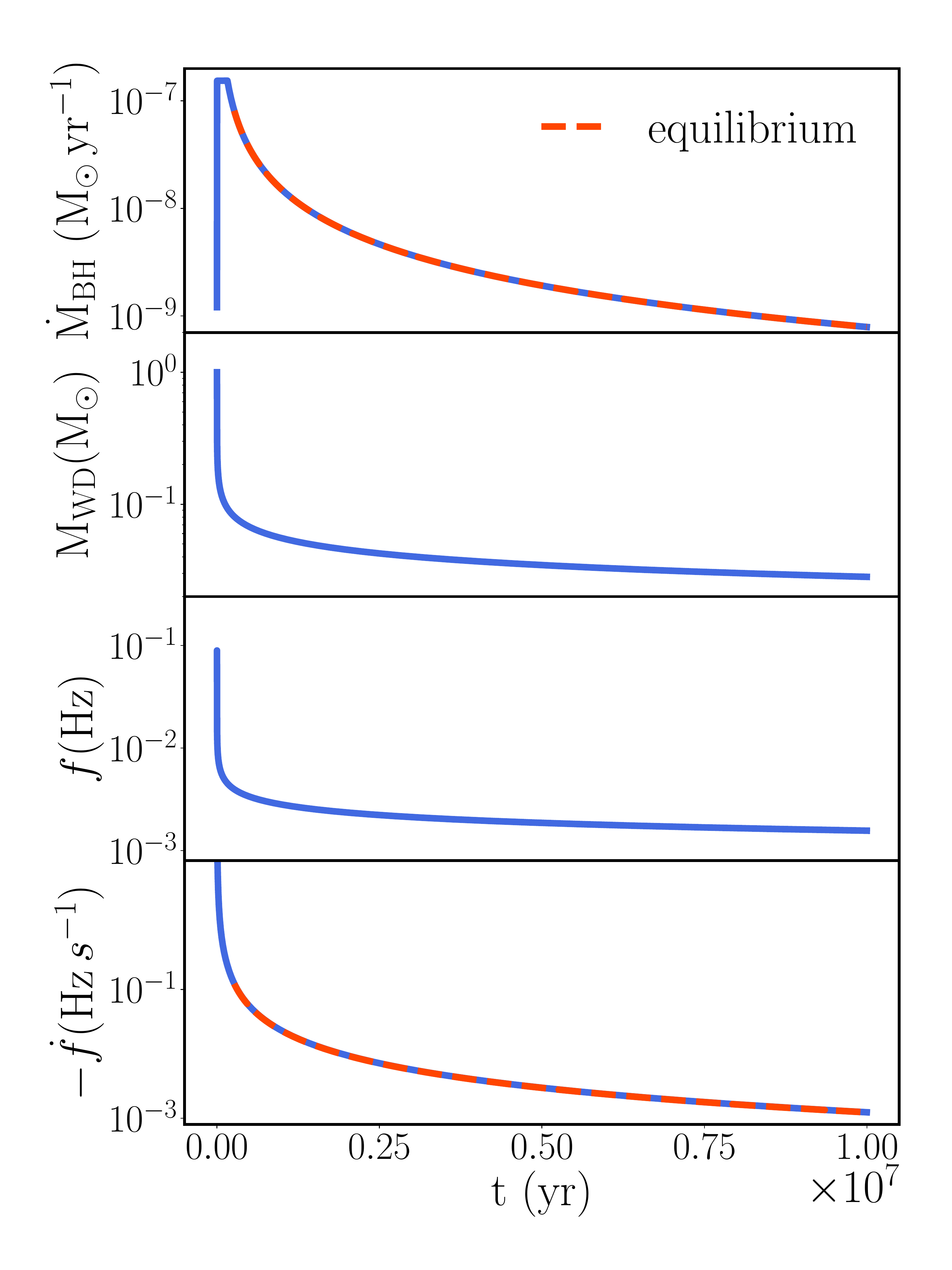}
	\caption{Evolution of the mass accretion rate, WD mass, GW frequency and its first derivative. The system has masses $ M_{\rm WD}= 1 \, M_{\odot}$ and $ M_{\rm BH}= 7 \, M_{\odot}$ at the time of first Roche lobe filling. The overlaid orange dashed line is the equilibrium solution described in App.~\ref{app:equil}.
	} \label{fig:single_evo}
\end{figure}

\subsection{Overfill and black hole spin}
We evolve the over-fill factor according to 
\begin{align}
\dot{\Delta} &= R_{\rm WD} \left[\left(\zeta_{\rm WD} - \zeta_{r_L} \right)\frac{\dot{M}_{\rm WD}}{M_{\rm WD}}  - \frac{\dot{a}}{a} \right], 
\label{eq:delta1}
\end{align}
where $ \zeta_{\rm WD} = \frac{{\rm d} \log R_{\rm WD}}{{\rm d} \log  M_{\rm WD}} $ and $ \zeta_{r_L} = \frac{{\rm d} \log R_L/a}{{\rm d} \log  M_{\rm WD}} $ can be derived using Eggleton's approximation for the mass-radius relationship of cold WDs and Eggleton's Roche lobe fitting formula~\citep{Eggleton}, respectively.

\begin{figure*}
	\centering 
	\subfloat{%
		\includegraphics[width=0.95\columnwidth]{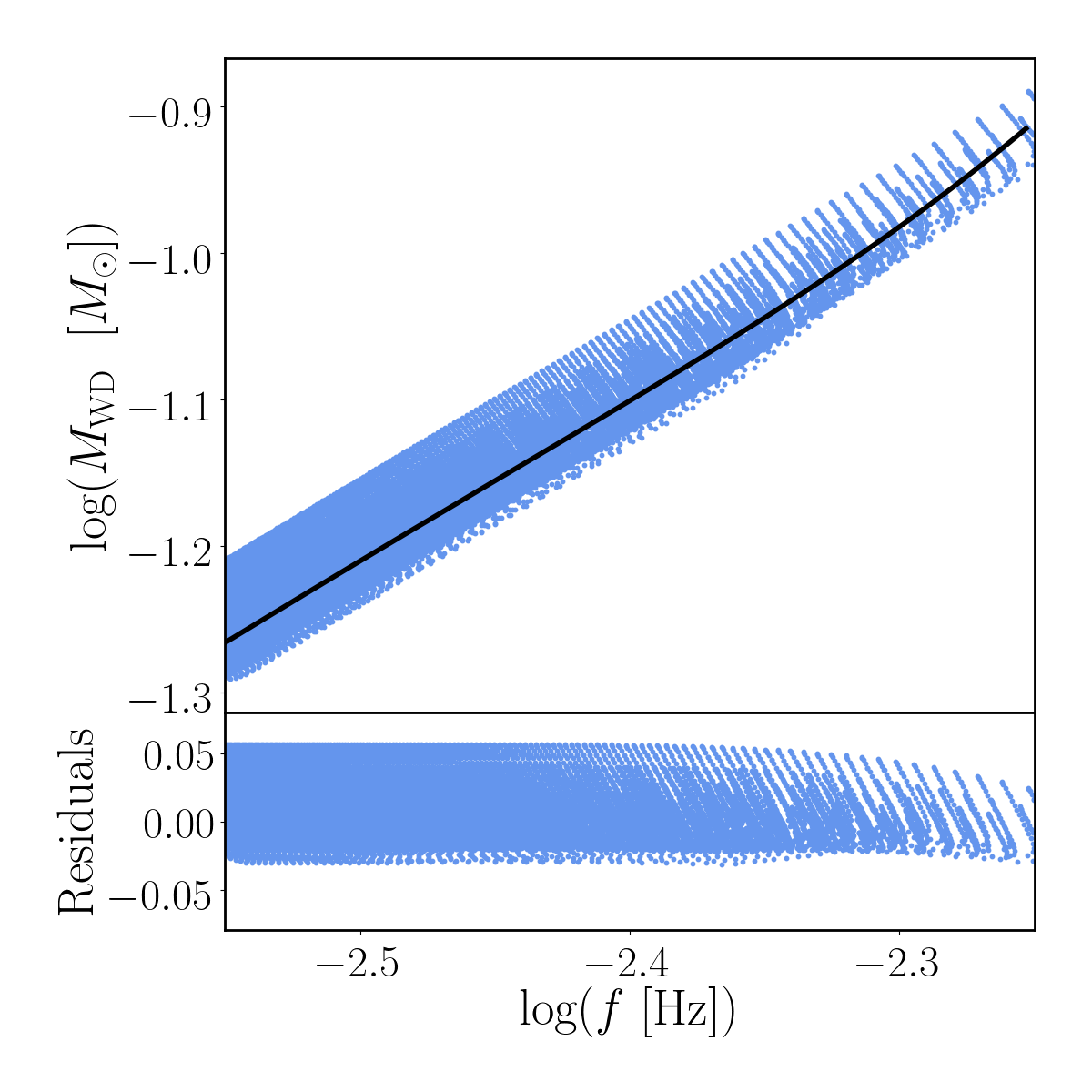}%
	}\qquad
	\subfloat{%
		\includegraphics[width=0.95\columnwidth]{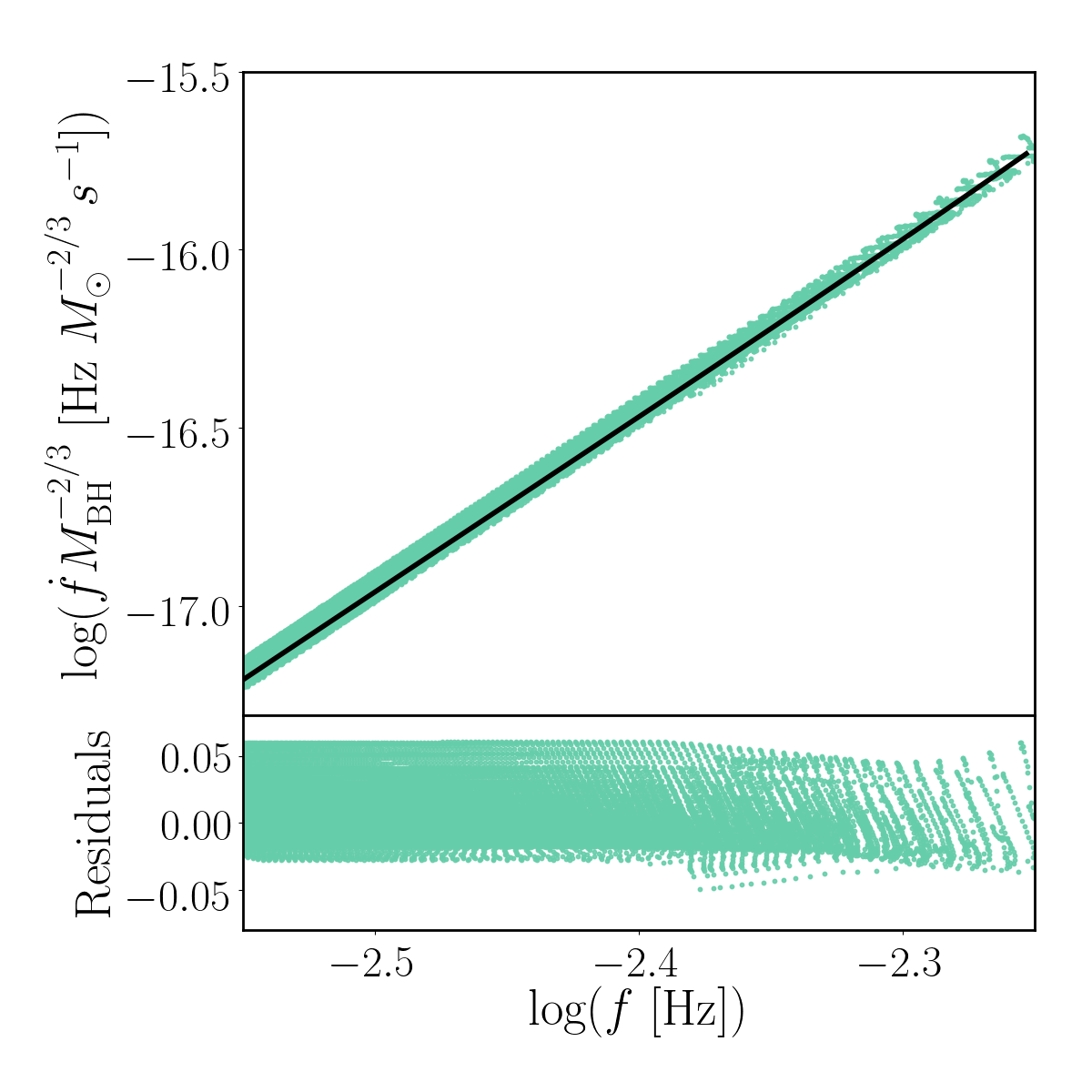}%
	}
\caption{ Evolutionary tracks of 400 WDBH binaries and their polynomial fits (black line). We focus on frequencies relevant to LISA. 
} \label{fig:evotr}
\end{figure*}

The angular momentum of the BH can be written in terms of the dimensionless spin $ \chi $,
\begin{align}
J_{\rm BH}=\frac{G}{c}M_{\rm BH}^2 \chi \label{j_1} \, .
\end{align}
The accreting BH will spin up according to Eq.~\eqref{eq:spin1}, from which we obtain
\begin{equation}\label{eq:spineq}
\dot{\chi}=\left ( \frac{c}{G} \frac{j_{\rm ISCO}}{M_{\rm BH}}  -2\chi \right )\frac{\dot{M}_{\rm BH}}{M_{\rm BH}} \, .
\end{equation}
The evolution of the BH spin is not our main focus and has little effect on the overall evolution of the binary. We therefore neglect for simplicity other factors affecting the spin evolution, such as radiation emitted by the accretion disk and fix the initial BH spin to $  \chi = 0.1 $. 

\subsection{Results}
We numerically integrate equations~\eqref{eq:M2eq}, \eqref{eq:M1eq}, \eqref{ev_jorb}, \eqref{eq:delta1} and \eqref{eq:spineq}, starting from the onset of mass transfer.
The long term evolution of a typical WDBH binary is shown in Figure~\ref{fig:single_evo}. The cap in the BH accretion rate on the top panel is due to accretion reaching the Eddington limit. As expected for mass-transfer dominated systems where the accretor is much more massive than the donor, the binary outspirals, giving a negative $ \dot{f} $.

Mass transfer proceeds rapidly at first, but quickly settles into an equilibrium rate. Equilibrium is attained when the increase in the Roche lobe matches the one in the WD radius. Thus, we obtain the equilibrium mass transfer rate by setting the right hand side (rhs) of Eq. \eqref{eq:delta1} to 0, see App. \ref{app:equil}.


\section{Universal relations}\label{univ_relation}

Across parameter space, the mass of the WD follows an evolutionary track as a function of the GW frequency, which is approximately independent of the accretor mass and the initial conditions, as displayed in Figure~\ref{fig:evotr}, left panel. 
We span initial WD masses between $[0.2, 1.2] \, M_{\odot}$, initial BH masses in the range $[3, 20] \, M_{\odot}$ and only keep points from the equilibrium stage. 
These tracks can be compared with the ones traced by WD accreting binaries in ~\cite{Breivik_2018}. Our WDBH tracks follow a slightly different trajectory and show a more pronounced dependence on the accretor mass, resulting in a larger spread in the tracks (and hence fit residuals). 

The absence of tidal interactions yields an additional relation between $\dot{f}M_{\rm BH}^{-2/3}$ and $f$. We show this relation in Figure~\ref{fig:evotr}, right panel. Once again, the relation is roughly independent of the accretor mass and initial conditions.
In App.~\ref{app:equil} we explain how this relation can be derived from the equilibrium solution.  

We fit both evolutionary track relations with a quartic polynomial $ \log(y) = \sum_{i=0}^{n} a_i \log(f [Hz])^i $ (Figure~\ref{fig:evotr}). The fit coefficients are listed in App.~\ref{app:fit}.

\section{Parameter estimation with LISA}\label{sec:LISA}

\begin{table*}
	\begin{center}
		\begin{tabular}{*{8}{c}}
			
			\cline{3-8}
			
			& & \multicolumn{3}{|c|}{\emph{HF}} &  \multicolumn{3}{c|}{\emph{LF}}  \\
			
			\cline{3-8}
			
			&  & \multicolumn{1}{|c}{$\tilde{M}_{\rm BH}$} & \multicolumn{1}{|c}{$\tilde{M}_{\rm WD}$} & \multicolumn{1}{|c}{$\tilde{D}_L$} & \multicolumn{1}{|c}{$\tilde{M}_{\rm BH}$} & \multicolumn{1}{|c}{$\tilde{M}_{\rm WD}$} & \multicolumn{1}{|c|}{$\tilde{D}_L$} \\
			\hline
			
			\multicolumn{1}{|c|}{\multirow{2}{*}{$100\%$}} & \multicolumn{1}{c|}{Fit}  & $0.99^{+0.01}_{-0.01}$ & $0.99^{+5.7 \times 10^{-8}}_{-5.8\times 10^{-8}}$ & $1.05^{+0.11}_{-0.14}$ & $1.04^{+0.40}_{-0.36}$ & $0.97^{+2.8\times 10^{-7}}_{-2.8\times 10^{-7}}$& \multicolumn{1}{c|}{$0.98^{+0.35}_{-0.29}$} \\
			\cline{2-8}
			
			\multicolumn{1}{|c|}{} & \multicolumn{1}{c|}{Full}  & $1.01^{+0.08}_{-0.04}$ & $0.99^{+0.02}_{-0.04}$ & $1.06^{+0.11}_{-0.15}$ & $1.03{+0.44}_{-0.38}$ & $0.98^{+0.04}_{-0.05}$ & \multicolumn{1}{c|}{$0.98^{+0.35}_{-0.29}$ }  \\
			
			\hline

			\multicolumn{1}{|c|}{\multirow{2}{*}{$75\%$}} &  \multicolumn{1}{c|}{Fit}    &  $0.99^{+0.01}_{-0.01}$ & $0.99^{+7.7 \times 10^{-8}}_{-7.8 \times 10^{-8}}$ & $1.39^{+0.16}_{-0.20}$ & $1.05^{+0.55}_{-0.47}$ & $0.97^{+3.8\times 10^{-7}}_{-3.8\times 10^{-7}}$& \multicolumn{1}{c|}{$1.28^{+0.59}_{-0.47}$}  \\
			
			\cline{2-8}
			
			\multicolumn{1}{|c|}{} &  \multicolumn{1}{c|}{Full}  &  $1.01^{+0.08}_{-0.04}$ & $0.99^{+0.02}_{-0.04}$ & $1.05^{+0.13}_{-0.16}$ & $1.03^{+0.61}_{-0.49}$ & $0.98^{+0.05}_{-0.05}$ & \multicolumn{1}{c|}{$0.96^{+0.44}_{-0.35}$}  \\
			
			\hline
			
		\end{tabular}
	\end{center}
	\caption{{ Uncertainties} on individual masses and distances normalized to the injected values, obtained with the fit to the global evolutionary tracks relations (Fit) and with the full results of numerical simulations (Full). The GW frequency $f$ and $\dot{f}$ are measured within $5 \times10^{-7} \ {\rm Hz}$ and $5 \times10^{-18}\ {\rm Hz} \ {\rm s}^{-1}$ for the \emph{HF} system, assuming a duty cycle of $75\%$. These measurements are an order of magnitude worse for the \emph{LF} system.}\label{errors}
\end{table*}

In the case of almost monochromatic sources such as WDBH and double WD binaries, the two GW polarizations take the simple form:
\begin{align}
h_{+}&=A_0 \frac{1}{2} \left (1+\cos^2(\iota) \right ) \cos(\phi_0+2\pi ft+\pi \dot{f} t^2), \\
h_{\times}&=A_0 \cos(\iota) \sin(\phi_0+2\pi ft+\pi \dot{f} t^2),
\end{align}
where $A_0=\frac{\mathcal{M}_c}{D_L}(\pi \mathcal{M}_cf)^{2/3}$ { is} the amplitude of the signal, $\mathcal{M}_c=M_{\rm BH}^3 M_{\rm WD}^3/M$ is the chirp mass of the binary, $\iota$ { is} the inclination of the binary with respect to the line of sight, and $\phi_0$ { is} the initial phase. Thus, GW observations provide us $A_0$, $f$ and $\dot{f}$ and we cannot infer the individual masses without further assumptions. 
In order to assess how the universal relations we derived can be combined with LISA measurements, we consider an accreting WDBH system at two different stages of its evolution: 
\begin{itemize}
 \item ``high frequency'' (\emph{HF}): $M_{\rm BH}=7.02 M_{\odot}$, $M_{\rm WD}=0.10 \ M_{\odot}$, 
 $f=5 \ {\rm mHz}$, $\dot{f}=-3.8 \times 10^{-16} \ {\rm Hz \, s^{-1}}$;
  \item ``low frequency'' (\emph{LF}): $M_{\rm BH}=7.02 M_{\odot}$, $M_{\rm WD}=0.06 \ M_{\odot}$, 
   $f=3 \ {\rm mHz}$, $\dot{f}=-3.2 \times 10^{-17} \ {\rm Hz \, s^{-1}}$; 
\end{itemize}
We compute LISA's response following~\cite{Cornish:2007if} to generate mock data and perform a full Bayesian analysis to infer the posterior distribution of the parameters of the source. For the noise level, we use the SciRdv1 curve~\citep{scirdv1} including a confusion noise due to the galactic foreground in addition to the instrument noise \cite{Mangiagli:2020rwz}. The parameter estimation is performed with the nested sampling algorithm \texttt{Multinest}~\citep{Feroz:2008xx}. We assume a mission duration of 6 years and two values of the duty cycle: $ 100\%$ and $75\%$. We set the distance to $D_L=10 \ {\rm kpc }$ and simulate the effect of a reduced duty cycle by placing the source further. For almost monochromatic sources, the angles essentially affect the signal to noise ratio (SNR) and have little impact on our analysis. 
For a duty cycle of $100\%$, the \emph{HF} and \emph{LF} systems have SNRs of $91$ and $26$ respectively. Systems at frequencies below $3\ {\rm mHz}$, although more numerous, have little chance of being detected due to the galactic foreground. With a duty cycle of $75\%$, $f$ and $\dot{f}$ are measured within $5 \times10^{-7} \ {\rm Hz}$ and $5 \times10^{-18}\ {\rm Hz.s^{-1}}$ for the \emph{HF} system and an order of magnitude worse for the \emph{LF} system.

In Table~\ref{errors} we report the estimates of the binary masses (normalized to the injected values) directly using the fits to the evolutionary tracks of Figure~\ref{fig:evotr}. We can use these results to infer the chirp mass and, from the measurement of $A_0$, the distance to the source. We find a reasonable agreement with the injected values (within $5 \%$). However for the \emph{HF} system, the injected values lie outside the $90\%$ confidence intervals. This is because the systematics of the model dominate over the statistical uncertainty. In particular, the very narrow range for $M_{\rm WD}$ is due to the extremely good measurement of $f$. 
To correct for this, we { estimate numerically} the values of $\alpha_1$ and $\alpha_2$ that best align the evolutionary tracks, $M_{\rm WD} \, M_{\rm BH}^{-\alpha_1}$ and $\dot{f}_{\rm WD} \, M_{\rm BH}^{-2/3-\alpha_2}$ as functions of $f$. The exponents $\alpha_1$ and $\alpha_2$ are frequency dependent and are determined for each system in the frequency range of observation. We then convolve LISA posteriors with the aligned tracks to infer $M_{\rm BH}$ and $M_{\rm WD}$. 

\begin{figure}[t]
	\centering
		\includegraphics[width=\linewidth]{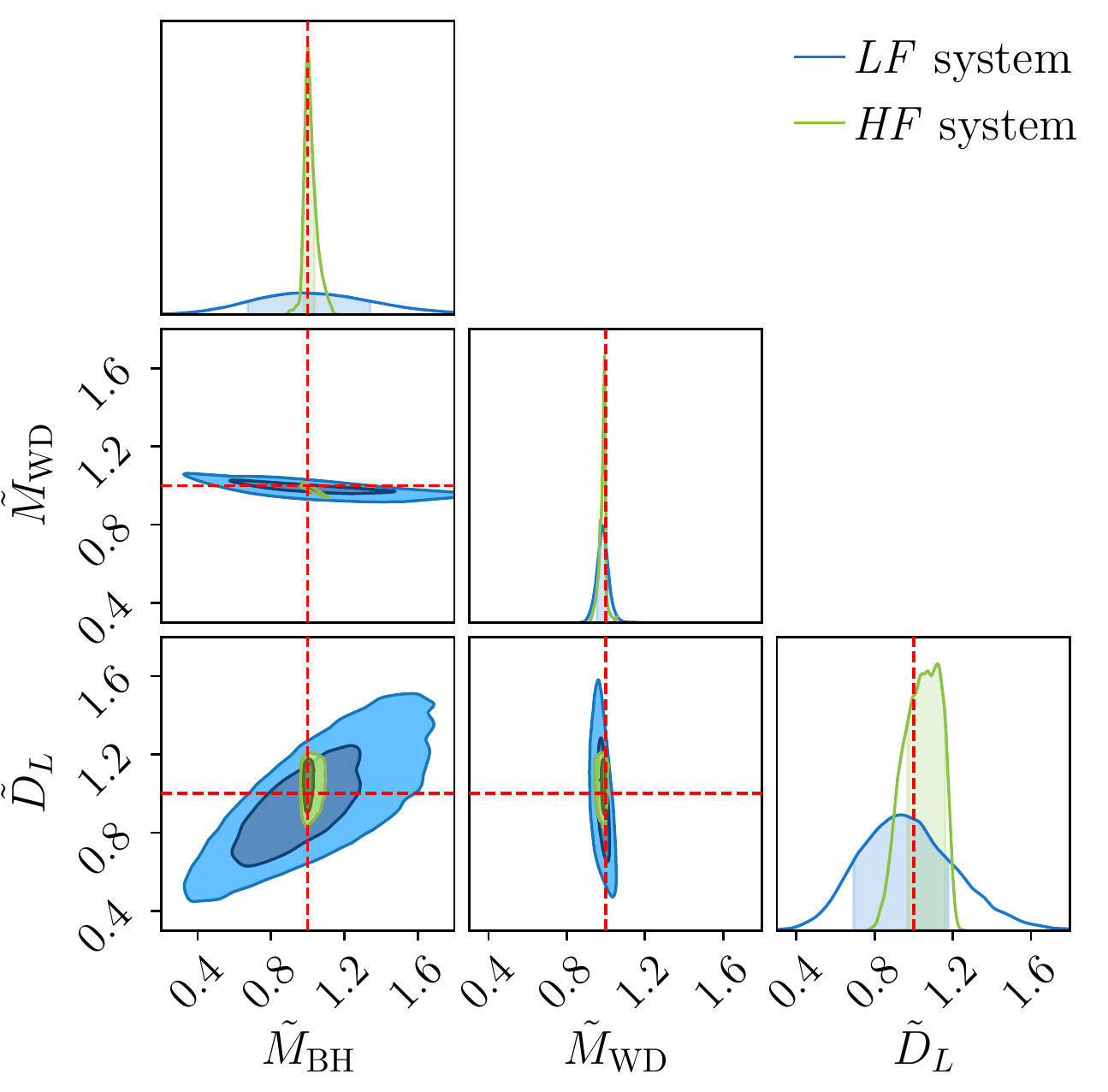} 
	\caption{ Posterior distributions for binary masses and luminosity distance normalized to the injected values for the high and low frequency systems, at $75\%$ 
		duty cycle. The contours indicate the 50 and 90 $\%$ confidence intervals and the dashed lines represent the true values (equal to $ 1 $ in our normalization). Posteriors are obtained with the rescaled universal relations, as described in Sec.~\ref{sec:LISA}. 
	} \label{fig:posteriors}
\end{figure}

In Figure~\ref{fig:posteriors}, we show how the measurement of $f$ and $\dot{f}$ together with this procedure translates into a measurement of the WD and BH masses for the two systems assuming a $100\%$ duty cycle. Table.~\ref{errors} also shows the improvement as compared to fit-based measurements and the very good agreement between the injected and the inferred values obtained with this procedure. $M_{\rm BH}$ is less well constrained than $M_{\rm WD}$ because it relies on the measurement of $\dot{f}$. The measurement is { worse} for the \emph{LF} system due to the lower value of $\dot{f}$ which results in it being measured not as well during the $6$ year mission. We note that the results are less affected by a reduced duty cycle. Finally, even in the worst scenario the { uncertainty} on $M_{\rm BH}$ is sufficiently small to unambiguously identify the accretor as a BH.  

\section{Conclusion}

Mass-transferring binaries containing a BH and a WD have been an elusive target, despite being predicted by population synthesis models. In this work we show that combining LISA observations with semi-analytic evolution models provides an estimate of the masses of both binary components as well as the distance to the source, { which is} information not usually accessible from galactic binary GW observations. 

WDBH binaries are potential sources of electromagnetic radiation, in particular X-ray emission. The \emph{HF} and \emph{LF} systems considered in this work would have respectively X-ray luminosity of $ 9 \, \times \, 10^{38} \, {\rm erg \, s^{-1}}$ and $ 1 \, \times \, 10^{38} \, {\rm erg \, s^{-1}}$ for radiative efficiency of $ 0.1 $, well within the capabilities of current facilities. The fact that we are yet to convincingly identify WDBH binaries among X-ray sources could be explained by the lower rates of these systems, and the difficulty to classify the binary components from electromagnetic emission alone. GW observations such as the ones described in this work, on the other hand, could unequivocally identify the BH companion. The very good localization of the source by LISA, $\mathcal{O}(1 \, {\rm deg}^2)$, could then provide the opportunity to observe an electromagnetic counterpart. In future work, we will explore the potential synergy between LISA and future electromagnetic surveys (Athena+, Square Kilometer Array) and the detectability of both GW and electromagnetic emission in the Milky Way and nearby galaxies.

To detect and learn the most from these systems with LISA, more detailed modeling will be crucial. This work did not take into account, for instance, the potentially disruptive effect of accretion winds on the accretion stream itself, and its potential variability on short timescales. 

Assuming a simple black body law for the BH and the WD, we estimate that emission from the BH disk could heat up the WD to $\mathcal{O}(10^5) \ {\rm K}$. Such low temperatures, if interpreted as core temperatures, have very little impact on the mass-radius relation~\citep{2020arXiv200807469B} and therefore the cold WD assumption remains a good approximation. 
A caveat is that the results of~\cite{2020arXiv200807469B} were obtained for cooling sequences of isolated WDs. 
More simulations of heated WDs as in~\cite{Piro:2005rd} could provide further insight on the effect of illumination on the evolution of WDBH binaries. WDs with masses lower then the ones considered in this work might also exhibit a stronger dependence on the temperature, see e.g.~\cite{Deloye_2003}.

Finally, we checked that the presence of tidal torques would not affect our results significantly for small synchronization timescales ($\tau \lesssim 100 \ {\rm yrs}$). This study could therefore apply to broader classes of galactic binaries.

\section*{Acknowledgements}
We are grateful to Katelyn Breivik and Gijs Nelemans for their constructive suggestions. A.T. is thankful to Stas Babak and Enrico Barausse for fruitful dicsussion and their support during the preparation of this work.
M.C.M.~thanks the Radboud Excellence Initiative for supporting his stay at Radboud University during part of this work. M.C.M.~and A.T.~are grateful for the hospitality of Perimeter Institute where part of this work was carried out. Research at Perimeter Institute is supported in part by the Government of Canada through the Department of Innovation, Science and Economic Development Canada, and by the Province of Ontario through the Ministry of Colleges and Universities. A.T.~was supported by the European Union's Horizon 2020 research and innovation program under the Marie Sk\l{}odowska-Curie grant agreement No 690904. A.T also acknowledges networking support from the COST Action CA16104.


\appendix

\section{Separation equation} \label{app:sep}

The equation for the orbital separation of the binary can be derived from Eqs.~\eqref{ev_jorb},~\eqref{eq:donorJ} and~\eqref{eq:spin1} and reads
	\begin{align}  
	\frac{\dot{a}}{2a}=-\frac{1}{1-3(1+q)kr_2^2}\Bigg[&
	\frac{\dot{J}_{\rm GW}}{J_{\rm orb}}
	+ \left( 1-\frac{q}{2(1+q)} + \frac{1}{2} q k r_2^2 +(1+q)\lambda k r_2^2 \right ) \frac{\dot{M}_{\rm WD}}{M_{\rm WD}} \nonumber \\
	&+ \left( q-\frac{q}{2(1+q)} + \frac{1}{2} q k r_2^2 + j_{\rm GR}\sqrt{(1+q)r_{\rm ISCO}}   \right ) \frac{\dot{M}_{\rm BH}}{M_{\rm WD}}
	\Bigg] \label{ev_sep}  ,
	\end{align}
where $r_i=R_i/a$.

\section{Equilibrium solutions} \label{app:equil}

After the initial phase of mass accretion, a very good approximation of the mass transfer rate can be obtained by setting the rhs of Eq.~\eqref{eq:delta1} to 0~\citep{Marsh:2003rd} and $ \dot{M}_{\rm BH} = - \epsilon_{\rm ISCO}\dot{M}_{\rm WD}$. We find
\begin{align}
\frac{\dot{M}_{\rm WD, \, e}}{M_{\rm WD}}=- \frac{\dot{J}_{\rm GW}/J_{\rm orb}}{K_{\rm eq}}  \label{eq:equil} ,
\end{align}
where
\begin{align}
K_{\rm eq}=&\frac{\zeta_{\rm WD} - \zeta_{r_L}}{2}(1-3(1+q)kr_{\rm WD}^2) +(1+q)\lambda k r_{\rm WD}^2(1-\epsilon_{\rm ISCO})(\frac{1}{2} q k r_{\rm WD}^2-\frac{q}{2(1+q)})+1\nonumber \\ &-\epsilon_{\rm ISCO}(q +j_{\rm GR}\sqrt{(1+q)r_{\rm ISCO}}) . \label{eq_ge}
\end{align}

Using Kepler's law, $\frac{\dot{f}}{2f}=-\frac{3\dot{a}}{2a}$ and replacing Eq.~\eqref{eq:equil} in Eq.~\eqref{ev_sep} gives, at equilibrium:
\begin{align}
 \frac{\dot{f}}{2f}= \frac{3 \dot{J}_{\rm GW}/J_{\rm orb} }{K_{\rm eq}(1-3(1+q)kr_2^2)}\Bigg[&1+ \left( q-\frac{q}{2(1+q)} + \frac{1}{2} q k r_2^2 + j_{\rm GR}\sqrt{(1+q)r_{\rm ISCO}}   \right ) \epsilon_{\rm ISCO} \nonumber \\
&-\left( 1+\frac{q}{2(1+q)} + \frac{1}{2} q k r_2^2 +(1+q)\lambda k r_2^2 \right ) \Bigg]. \label{fdot_f}
\end{align}
Furthermore, 
\begin{equation}
\frac{\dot{J}_{\rm GW}}{J_{\rm orb}} = \frac{32}{5} \frac{G^3}{c^5} \frac{M_{\rm BH} M_{\rm WD} M}{a^4} \propto M_{\rm BH} M_{\rm WD} M \left (  \frac{M}{f^2} \right )^{4/3} \simeq M_{\rm BH}^{2/3} M_{\rm WD} f^{8/3}.
\end{equation}
where in the last step we used $M_{\rm WD} \ll M_{\rm BH}$, so that $M \simeq M_{\rm BH}$. 
Finally, the late time evolution of the other terms in Eq.~\eqref{fdot_f} happens to have a weak dependence on $M_{\rm BH}$, so $\dot{f} M_{\rm BH}^{-2/3}$ is an almost $M_{\rm BH}$ independent quantity as verified in Fig. \ref{fig:evotr}, right panel.

\section{Fits to the evolutionary tracks}\label{app:fit}
The coefficients for the evolutionary tracks fits described in the main text are summarized in Table~\ref{tab:fit}.

\begin{table*}[h!]
	\begin{center}
		\begin{tabular}{*{6}{c|}}
						
			\cline{2-6}
			
			& $a_0$ & $a_1$ & $a_2$ & $a_3$ & $a_4$  \\
			
			\hline
			
			\multicolumn{1}{|c|}{\emph{$y = M_{\rm WD} [M_{\odot}] $}} & 319.7593186
			& 509.0101135
			& 303.8011829
			& 80.7077869
			& 8.0347503
			\\
			\hline
				\multicolumn{1}{|c|}{\emph{$y = \dot{f} M_{\rm BH}^{-2/3} [ {\rm Hz} \, M_{\odot}^{-2/3} s^{-1}] $}} & 142.6384491
				& 236.4026829
				& 136.2183828
				& 35.5719325
				& 3.4778346
			\\
			\hline
		\end{tabular}
	\end{center}
	\caption{Coefficients for the fits in Figure~\ref{fig:evotr}.}\label{tab:fit}
\end{table*}

%


\end{document}